\documentclass[pra,superscriptaddress,preprint]{revtex4}
\usepackage{blindtext}
\usepackage{amsfonts}
\usepackage{amsmath,amssymb}

\usepackage{graphicx}
\usepackage{xcolor}
\usepackage{subfigure}
\usepackage{bm}
\usepackage{epstopdf}
\begin{document}
\title{Pairing and molecule formation along the BEC-BCS crossover for finite range potentials}
\author{E. Neri}
\author{S. F. Caballero-Ben\'itez}
\author{V. Romero-Roch\'{\i}n}
\author{R. Paredes}
\email{rosario@fisica.unam.mx}
\affiliation{Instituto de F\'{\i}sica, Universidad
Nacional Aut\'onoma de M\'exico, Apartado Postal 20-364, M\'exico D.
F. 01000, Mexico. }

%\pacs{67.85.-d, 03.75.Ss, 03.75.Gg}

\begin{abstract}
We analyze the BCS-BEC crossover transition of a balanced two component mixture of fermions interacting via a finite range potential, within a mean field approach. For the analysis we consider three finite range potentials cases describing the interaction between different Fermi species: a square well, an exponential and a Yukawa potential. The $T=0$ thermodynamics analysis along the BCS-BEC crossover allow us to recognize the proper variables, for finite range interactions, that capture the transition from a thermodynamic non-universal behavior at unitarity, to its universal restoration. On the other side, the determination of the pair functions along the crossover suggests that the smooth transition occurs always between the scattering resonance and the change of sign of the chemical potential. This identification follows directly from the pair wave functions behavior, which in the BCS and BEC sides become exponentially localized and oscillatory slowly decaying respectively. 
\end{abstract}

\maketitle

\section{Introduction}
The observation in 2004 of the transition from the atomic Bardeen-Cooper-Schriefer state to the molecular Bose-Einstein condensate, known as the BCS-BEC crossover \cite{Regal, Zwierlein, Kinast, Bourdel, Chin-Science}, confirmed the early prediction \cite{Blatt, Eagles, Leggett, Nozieres, Holland, Timmermans, Ohashi} that these two different states, in principle opposite in nature, emerge from the same origin, namely, from an interacting gas composed of a balanced mixture of Fermi atoms in two different hyperfine states. Even more, those experiments established that the crossover occurs due to the presence of a Feshbach resonance in very low energy two-body collisions. Recently, exotic phases of quantum matter exhibiting this smooth transition have been investigated in nuclear matter\cite{Strinati}, color superfluidity\cite{SadeMelo} and FFLO states \cite{Aalto} among an extensive amount of possible scenarios. Moreover, there have been experiments of quantum simulation with ultracold quantum matter exploring vortex dynamics\cite{Zwierlein2} and transport properties\cite{Esslinger,Roati1,Roati2}, which have allowed to study physical phenomena in the parameter regime of the crossover.

Based on the facts that, on the one side, the mean-field approach provides an intuitive and qualitatively reasonable description of the BCS-BEC crossover, and on the other, the two-channel Feshbach resonances used in real experiments can be replaced by a single-channel potential resonance, we investigate in this work the physics of the crossover for different finite range interaction potentials. In addition, we track the form of the pair functions along the whole crossover, as a function of the parameters that define the model potential.
 
We recall that, typically, the BCS-BEC crossover is studied by using the so-called contact approximation \cite{Leggett}, in which the finite range interaction potential is replaced by a Dirac delta potential proportional to the (positive) two-body collision scattering length. Since actual potentials are finite range, and do not depend explicitly on the scattering length, there has already been interest in analyzing the crossover and its properties taking those aspects into account. 
Previous analysis of the $T=0$ BCS-BEC problem considering finite range potentials include both homogeneous and inhomogeneous cases and are based on Quantum Monte Carlo approaches \cite{Jauregui, QMC1, QMC2}, as well as on numerical solutions of the coupled equations of the BCS theory \cite{Meera,Caballero-Benitez}. Here we show that the structure of the coupled equations for the energy gap and the density leads us to identify closed expressions for any finite range potential model with definite Fourier transform, thus, providing a route to address a variety of the physics of the crossover. According to the universality hypothesis,\cite{Ho} in the limit when the scattering length diverges and therefore the system losses its atomic length-scale, it is expected that all thermodynamic quantities become independent on any atomic length, including the characteristic range defining the interaction potential. At the same time the so called unitarity regime, namely the strongly interacting regime near the resonance, marks the separation of the physically quite different BCS and BEC limits. Since all the above interatomic potentials depend on two parameters, namely, its energy depth and its spatial range, and none of these diverge at unitarity, one should not expect an absolute universality even at the resonance. However, as we shall see, the present investigation indicates that, while universality is restored in the limit of very diluted gases, it also suggests that a more appropriate variable to define unitarity is a dimensionless quantity that depends on the two parameters of the potential, since such a variable appears explicitly in all thermodynamic variables. The present investigation also brings a new insight into the pairing phenomenon to explore the physics very diverse systems, for instance, fluids immersed in the ultracold atom scheme, such as a gas of cold atoms in optical cavities where tunable range effective atomic interactions can be engineered \cite{Mekhov1,Caballero-OC-1,Caballero-OC-2,Caballero-OC-3}, or systems belonging to the context of high energy physics as the relativistic description of BCS-BEC crossover in nuclear matter \cite{Sun,Toledo,Sun2}.

This manuscript is organized in 5 sections. In section 2 we write the BCS mean field equations that describe the $T=0$ ground state of the interacting Fermi mixture for the finite range potentials here analyzed. In section 3 we present the thermodynamics along the BCS-BEC crossover as a function of both, the $s-$wave scattering length and the parameters $R$ and $V_0$ associated to the range and depth of the model potential. In section 4 the analysis of the pair functions is presented. Finally a discussion and a summary of this work is presented in section 5.  

\section{BCS theory for finite range potentials}
The most general Hamiltonian describing a balanced mixture of Fermi atoms interacting between pairs, in the grand canonical ensemble is, 

\begin{eqnarray}
 H-\mu N &=& \sum_{\mathbf{k}, \sigma} \hat n_{\mathbf{k}, (\sigma}\epsilon_{\mathbf{k}}-\mu) + \nonumber \\
&& \frac{1}{2}\sum_{\mathbf{K};\,\sigma,\sigma'} U_\mathbf{k k'}\hat{a}_{\mathbf{k+ q},\sigma}^{\dagger}\hat{a}_{\mathbf{k'-q},\sigma'}^{\dagger}\hat{a}_{\mathbf{k'},\sigma'}\hat{a}_{\mathbf{k},\sigma},\label{eq:hamiltonian}
\end{eqnarray}
with $\epsilon_{\mathbf{k}}= \frac{\hbar^2 \mathbf{k}^2}{2m}$, $m$ being the atom mass, $\mathbf{K}= \{ \mathbf{k}, \mathbf{k}',\mathbf{q} \}$ and  $\hat n_{\mathbf{k}, \sigma}= \hat{a}_{\mathbf{k},\sigma}^{\dagger}\hat{a}_{\mathbf{k},\sigma}$ the number operator written in terms of the creation and annihilation Fermi operators $\hat{a}_{\mathbf{k},\sigma}^{\dagger}$ and $\hat{a}_{\mathbf{k},\sigma}$ respectively. $\sigma$ labels two different hyperfine spin states that we shall denote as $\uparrow$ and $\downarrow$. $U_{\mathbf{kk'}}$ is the Fourier transform of the finite range potential that models the inter particle interaction $U_{ \mathbf{k} \mathbf{k}'} = \int  e^{i (\mathbf{k} - \mathbf{k}') \cdot \mathbf{r} }U(r) d^3r$. As it is well known, consideration of momentum conservation leads us to recognize three different terms associated to the interaction between pairs, they are, 

\begin{equation}
\begin{aligned}
H_{H} & ={\displaystyle \frac{U_{0}}{2V}\sum_{\mathbf{k} \mathbf{k}'}}\,\left(\hat n_{\mathbf{k}\uparrow}+\hat n_{\mathbf{k}\downarrow}\right)\left(\hat n_{\mathbf{k}'\uparrow}+\hat n_{\mathbf{k}'\downarrow}\right),\\
H_{F}&=-\frac{1}{2V}\sum_{\mathbf{k},\mathbf{k}'}\,U_{\mathbf{k} \mathbf{k}'}\left(\hat n_{\mathbf{k}\uparrow}\hat n_{\mathbf{k}'\uparrow}+\hat n_{\mathbf{k} \downarrow}\hat n_{\mathbf{k}' \downarrow}\right),\\
H_{p} &={\displaystyle \frac{1}{V}\sum_{\mathbf{k} \mathbf{k}'}}\,U_{\mathbf{k} \mathbf{k}'}\,\hat a_{\mathbf{k}'\uparrow}^{\dagger} \hat a_{-\mathbf{k}'\downarrow}^{\dagger}\hat a_{\mathbf{k}\uparrow}\hat a_{-\mathbf{k}\downarrow},\\
\end{aligned}
\end{equation}
where $U_0= U_{\mathbf{k} \mathbf{k}'}(0)$. In the standard BCS-BEC crossover theory, the term of pairs $H_{p}$ is the responsible of the paring phenomenon, while the terms associated to Hartree $H_{H}$ and Fock $H_{F}$ contributions are typically neglected. Strictly speaking, in the case of the contact potential approximation, the terms $H_{H}$ and $H_{F}$ show a divergence when the $s$-wave scattering length diverges. As it is well known, such a divergence is an artifact associated to the contact approximation itself. As previously shown in the literature, those divergences do not appear for finite range potentials \cite{Caballero-Benitez}. In fact, for the finite range potentials here analyzed, the contributions $H_{H}$ and Fock $H_{F}$ do not exhibit any anomalous behavior along the BCS-BEC crossover. Following the formulation made by Leggett \cite{Leggett},  we shall neglect in our work these terms representing just an energy shift. Thus, after substituting the BCS ansatz  for the many-body ground state wavefunction, $|\Psi_{\mathrm{BCS}}\rangle= \Pi_{\mathbf{k}} \left(u_{ \mathbf{k}}+v_{ \mathbf{k}} a_{ {\mathbf{k}} \uparrow}^{\dagger} a_{- {\mathbf{k}} \downarrow }^{\dagger} \right) |0 \rangle )$, and performing the standard variational procedure, one obtains the grand potential function $\Omega= \langle \Psi_{\mathrm{BCS}}| H-\mu N| \Psi_{\mathrm{BCS}} \rangle$,
\begin{equation}
\Omega= \sum_{ \mathbf{k}} (\epsilon_{\mathbf{k}} - \mu) v_{\mathbf{k}} + \frac{1}{V} \sum_{\mathbf{k}, \mathbf{k}'} U_{\mathbf{k}\mathbf{k}'} u_{\mathbf{k}'} v_{\mathbf{k}'} u_{\mathbf{k}} v_{\mathbf{k}},\label{Omega}
\end{equation}
where $2u_{\mathbf{k}}^2=1+ (\epsilon_{\mathbf{k}}-\mu)/ E_{\mathbf{k}}$ and $2v_{\mathbf{k}}^2=1- (\epsilon_{\mathbf{k}}-\mu)/ E_{\mathbf{k}}$, with $E_{\mathbf{k}}=\sqrt{ (\epsilon_{\mathbf{k}}-\mu)^2+ \Delta_{\mathbf{k}}^2 }$, thus defining the $\mathbf{k}-$dependent energy gap $\Delta_{\mathbf{k}}$. The grand potential $\Omega$ is minimized when the energy gap and and the chemical potential $\mu$ satisfy the coupled equations,
\begin{equation}
\begin{aligned}\Delta_{\mathbf{k}} & =-\frac{1}{V}\sum_{\mathbf{k}}U_{\mathbf{k}\mathbf{k}'}\frac{\Delta_{\mathbf{k}'}}{2E_{\mathbf{k}'}}\,,\\
N & =\sum_{\mathbf{k}}\left(1-\frac{\epsilon_{\mathbf{k}} - \mu}{E_{\mathbf{k}} }\right).
\end{aligned}
\label{coupledeqs}
\end{equation}
where $N$ is the total number of atoms. In the thermodynamic limit, scaling the energies with respect to the Fermi energy $\epsilon_F= \hbar^2 k_{F}^2/2m$, with $k_F= (3 \pi^2 N/V)^{1/3}$ the Fermi wave vector, those coupled equations (\ref{coupledeqs}) can be expressed in a general form for an arbitrary central potential $U(r)$ as,

\begin{equation}
\begin{aligned}\tilde{\Delta}_{x} & =-\frac{1}{16\pi^{3}}\intop_{0}^{\infty}\tilde{F}\left(x,x'\right)\frac{\tilde{\Delta}_{x'}}{2\tilde{E}_{x'}}\,\sqrt{x'}\,dx'\,,\\
1 & =\frac{3}{4}\intop_{0}^{\infty}\left(1-\frac{x-\tilde {\mu}}{\tilde{E}_{x}}\right)\,\sqrt{x}\,dx\,,
\end{aligned}
\label{SE-Delta-N2}
\end{equation}
where $x=k^2/k_F^2$ is a dimensionless variable. Here and henceforth tilde refers to dimensionless variables. The function $\tilde{F}\left(x,x'\right)$, namely, the kernel of the energy gap equation, depends on the specific form of the potential that models the interaction $U\left(r \right)$. Written in spherical coordinates this function adopts the form: $\tilde F \left(x,x'\right)= \int \tilde U_{\mathbf{k} \mathbf{k}'}  \sin \theta d\theta d \phi  $. One can find a simplified expression for $\tilde{F}\left(x,x'\right)$ for an arbitrary spherical potential of the form $U\left(| {\mathbf{r}} \right|)=V_0 u \left( r/R \right)$, where $V_0$ and $R$ are the potential depth and the spatial range respectively, and $u \left( r/R \right)$ a dimensionless function that depends on the ratio $r/R$. The final form for $\tilde{F}\left(x,x'\right)$ is:

\begin{equation}
\tilde{F} \left( x, x, {\tilde R}, {\tilde V}_0 \right) =-\frac{8\pi^{2} \tilde{V_0} }{\sqrt{x\,x'}}\intop_{0}^{\infty}u\left( {\tilde r} /{\tilde R} \right)\left[\cos\left(y_{2}\right)-\cos\left(y_{1}\right)\right]\,d {\tilde r},
\end{equation}
where $y_1=\tilde{r} \,\left| \sqrt{x} -\sqrt{x'}  \right|$ and $y_2=\tilde{r} \,\left| \sqrt{x} + \sqrt{x'}  \right|$.
In this manuscript we shall consider three different cases of the potential modeling the interparticle interactions, a square well, an exponential, and the Yukawa potentials:

\begin{equation}
\begin{aligned}U_{\mathrm{sw}}\left(r\right) & =\begin{cases}
-V_{0} & r\leq R\\
0 & r\geq R,
\end{cases}\\
U_{\mathrm{exp}}\left(r\right) & =-V_{0}e^{-r/R},\\
U_{\mathrm{Yuk}}\left(r\right) & =-V_0 \frac{R}{r}e^{-r/R}.
\end{aligned}
\label{potentials}
\end{equation}
A natural dimensionless quantity that characterizes these potentials is $\chi = \left(2m R^2 V_0/\hbar^2\right)^{1/2}$. As we will see below, this quantity and the scaled potential range $\tilde R = k_F R$ are very appropriate to describe the overall role of the finite range potential, since the first one determines the relative values of $V_0$ and $R$ at the resonance, and the second one yields the degree of diluteness of the gas; we will be mainly interested in the case $\tilde R \ll 1$.

 The function $\tilde{F}$ for these potentials is given by,
\begin{equation}
\begin{aligned}
\tilde{F}_{\mathrm{sw}}\left(x,x'\right)&=-\frac{8\pi \tilde{V}_0 \tilde{R} }{\sqrt{x x'}}\, \left( \frac{ \sin Y_2 }{Y^2_2} - \frac{\sin Y_1}{Y^2_1} \right),\\
\tilde{F}_{\mathrm{exp}}\left(x,x'\right)&=-\frac{8\pi \tilde{V}_0 \tilde{R} }{\sqrt{x x'}}\, \left( \frac{1}{1+Y^2_2} - \frac{1}{1+Y^2_1} \right) ,\\
\tilde{F}_{\mathrm{Yuk}}\left(x,x'\right)&=-\frac{8\pi \tilde{V}_0 \tilde{R} }{\sqrt{x x'}}\, \mathrm{ln} \left( \frac{1+Y^2_2}{1+Y^2_1} \right)^{1/2}, 
\label{Fs}
\end{aligned}
\end{equation}
where now  $Y_1=\tilde{R} \,\left| \sqrt{x} -\sqrt{x'}  \right|$ and $Y_2=\tilde{R} \,\left| \sqrt{x} + \sqrt{x'}  \right|$.

Solution of the coupled system (\ref{SE-Delta-N2}) gives the full information of the crossover for the studied interacting potentials. As we shall see in the next section, this coupled system will be solved for a fixed density $N/V$ and the parameters that define the two-body interaction $V_0$ and $R$. Here we want to stress two different aspects exhibited by finite range potentials, with respect to its counterpart, the contact interaction case. The first one is that the energy gap $\tilde{\Delta}_{x}$ is not a constant quantity as in the standard BCS theory, instead, it is a function of the momentum $\mathbf{k}$. The second aspect to be emphasized is that, as a consequence of having a potential depending on the range $R$, it is expected that the physical properties capture such a dependency. In other words, one should expect that besides the natural Fermi length $1/k_F$ present in the usual contact interaction potential, the presence of the additional length become noticeable through the whole BCS-BEC crossover.

A technical but relevant observation that should be pointed out is that, in addition for the interatomic potential to exhibit a potential resonance, thus signaling the emergence of a bound state, the interatomic potential should preferably have an analytic Fourier transform, otherwise the numerical calculations can become prohibitively expensive and inaccurate. Certainly, the necessity of an analytic expression for the $s-$wave scattering length in terms of the parameters $V_0$ and $R$ is not an essential requirement, and to the best of our knowledge, just few potentials have analytic expressions for the $s-$wave scattering length. As a matter of fact, for the model potentials here analyzed only two of them have such a relationship; the other can be found numerically. This is discussed in the following section.

\subsection{$s$-wave scattering length}

In this subsection we present the results of the $s-$wave scattering length for the interacting potentials of our study, Eq. (\ref{potentials}), $U_{\mathrm{sw}}(r)$, $U_{\mathrm{exp}}(r)$ and $U_{\mathrm{Yuk}}(r)$. As it is known, while the square well and exponential potentials have an explicit relationship for the $s-$wave scattering length $a$ in terms of the parameters $V_0$ and $R$, \cite{Rarita,square} the Yukawa potential does not. 
\begin{figure}[t!]
\includegraphics[width=0.8\textwidth]{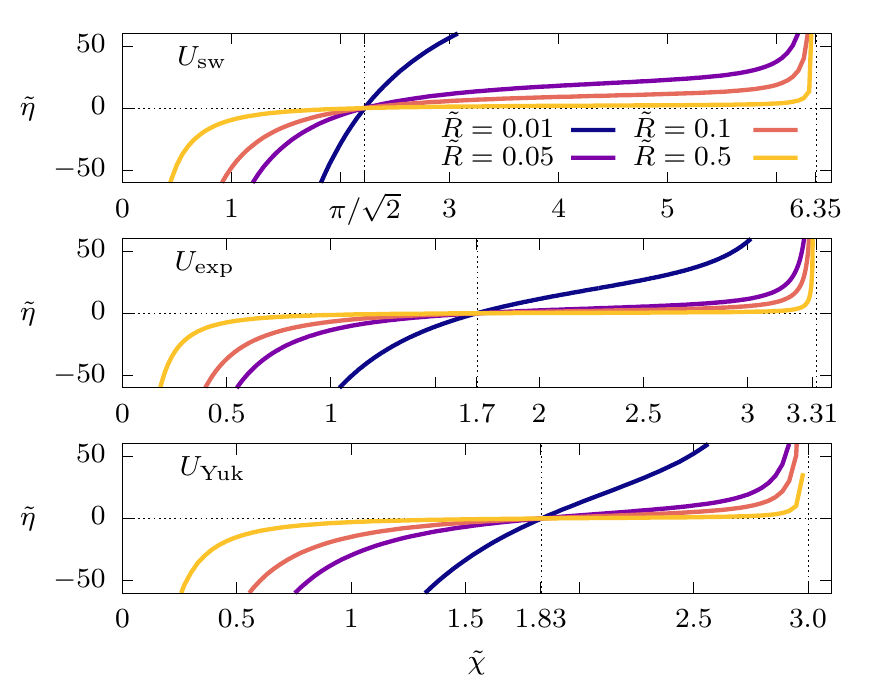} 
\caption{\footnotesize (Color online) Inverse of the $s-$wave scattering length $\tilde \eta = 1/k_F a$ as a function of the variable $\chi= \tilde R \sqrt{ \tilde V_0}$, for the square well $U_{\mathrm{sw}}(r)$, exponential $U_{\mathrm{exp}}(r)$, and Yukawa potentials $U_{\mathrm{Yuk}}(r)$, Eqs. (\ref{potentials}). Different curves in each panel correspond to different values of potential range $R$. The vertical dashed line indicates the value of $\chi$ for which the $s-$wave scattering length diverges in each case.}
\label{Fig1}
\end{figure}
The expressions for $a$ associated to each potential are,
\begin{equation}
\begin{aligned}
\tilde a_{\mathrm{sw}}(\chi)&=\tilde R \left[1- \frac{\tan \left ( \chi/ \sqrt{2}\right)}{\chi/ \sqrt{2}} \right ],\\
\tilde a_{\mathrm{exp}}(\chi)&= -2 \tilde R \left[ \frac{\pi}{2} \frac{N_0(\sqrt{2} \chi)}{J_0(\sqrt{2} \chi)} - \ln \left( \frac{\chi}{\sqrt{2}} \right ) - \gamma \right]
\label{scattering}
\end{aligned}
\end{equation}
where $\gamma$ is the Euler-Mascheroni constant, and $J_0(x)$ and $N_0(x)$ are zero order Bessel functions of first and second kind. These expressions for the $s-$wave have been written in its dimensionless form. In particular, the dimensionless parameter $\chi$ can also be expressed in terms of reduced quantities as $\chi = \tilde R\sqrt{ \tilde V_0}$. The scattering length for the Yukawa potential can be obtained numerically by solving the appropriate two-body problem, see Ref. \cite{Jeszenszki}. In Fig.\ref{Fig1} we plot the inverse of the dimensionless $s-$wave scattering length $\tilde \eta = 1/k_F a$ as a function of $\chi$ for several values of the range $\tilde R$ along the first potential resonance, for the three potentials considered. As one can see from this figure, for a given potential, the emergence of a bound state occurs at a fixed value of the variable $\chi$, which is where all the curves coalesce, regardless of the value $\tilde R$. In other words, at resonance, there is a unique or universal value of $\chi$. These values are $\chi = \pi/\sqrt{2}$, 1.7 and 1.83, the last two approximately, for the square well, exponential and Yukawa potentials, respectively.

According to the universal hypothesis at unitarity, which is when the $s-$wave scattering length of the two-body problem diverges, the thermodynamics of the gas becomes independent on the specific details of the two body interaction. Thus presumably, one should expect that differences of the physical properties become absent at least at $\tilde \eta=0$.  As we will see this is not quite true for arbitrary values of $\tilde R$, however, as $\tilde R \to 0$, namely, in the very dilute limit where the theory is ultimately expected to be valid, unitarity is restored.

\section{$T=0$ thermodynamics in the BCS-BEC crossover}

\begin{figure}[t!]
\includegraphics[width=0.9\textwidth]{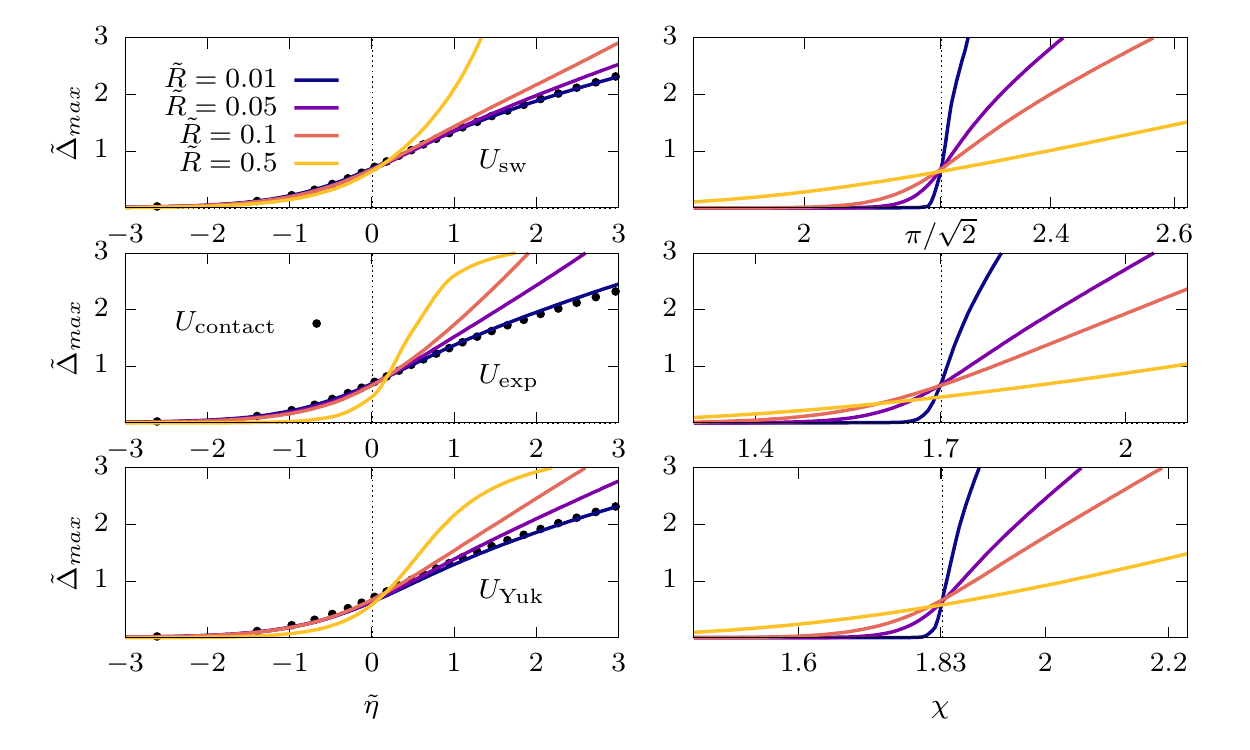} 
\caption{\footnotesize (Color online) Dimensionless energy gap (maximum) $\tilde \Delta_{\mathrm {max}}$as a function of the inverse of the $s-$wave scattering length $\tilde \eta$ (left panels) and the dimensionless variable $\chi$ (right panels), for the square well $U_{\mathrm{sw}}$, exponential $U_{\mathrm{exp}}$, and Yukawa potentials $U_{\mathrm{Yuk}}$. The dotted line in the left panels is the contact approximation result. Different curves in each panel correspond to different values of the potential range $\tilde R$. The vertical dotted line in the right panels indicate the values of $\chi$ at resonance for the different potentials.}
\label{gap}
\end{figure}

\begin{figure}[t!]
\includegraphics[width=0.9\textwidth]{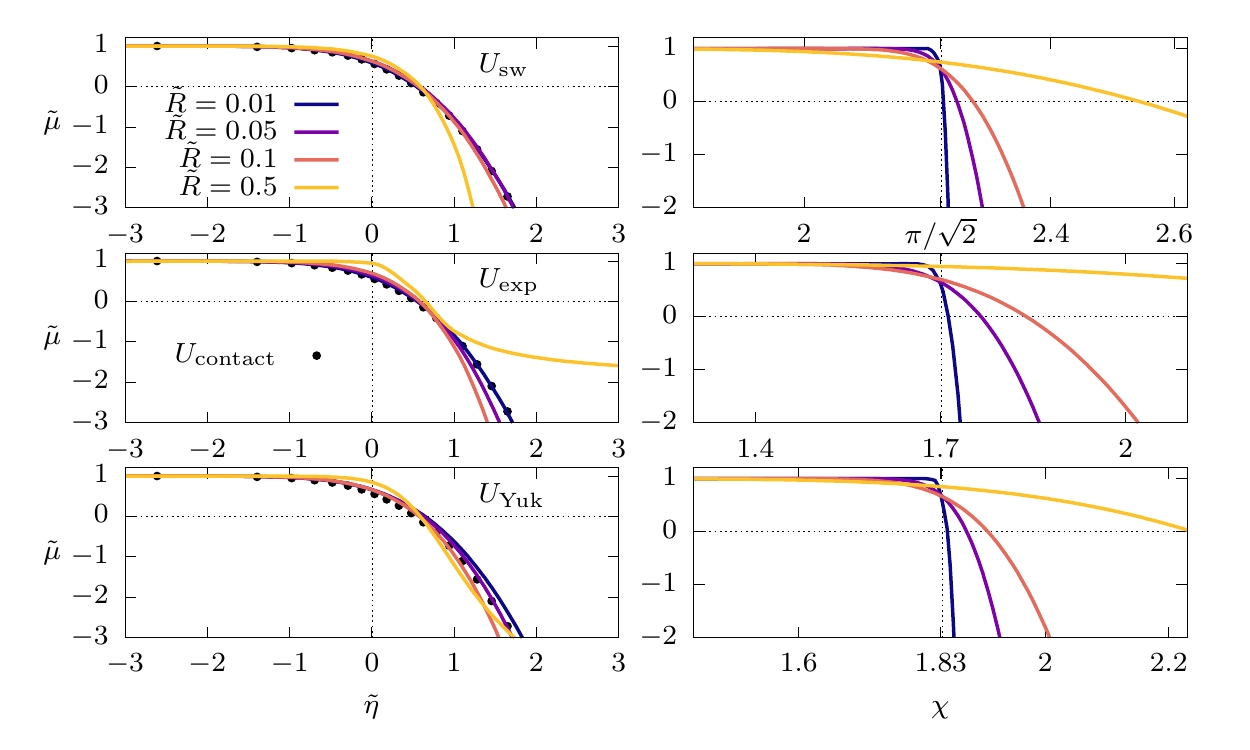} 
\caption{\footnotesize (Color online) Dimensionless chemical potential $\tilde \mu$ as a function of the inverse of the $s-$wave scattering length $\tilde \eta$ (left panels) and the dimensionless variable $\chi$ (right panels), for the square well $U_{\mathrm{sw}}$, exponential $U_{\mathrm{exp}}$, and Yukawa potentials $U_{\mathrm{Yuk}}$. The dotted line in the left panels is the contact approximation result. Different curves in each panel correspond to different values of the potential range $\tilde R$. The vertical dotted line in the right panels indicate the values of $\chi$ at resonance for the different potentials.}
\label{mu}
\end{figure}

\begin{figure}[t!]
\includegraphics[width=0.9\textwidth]{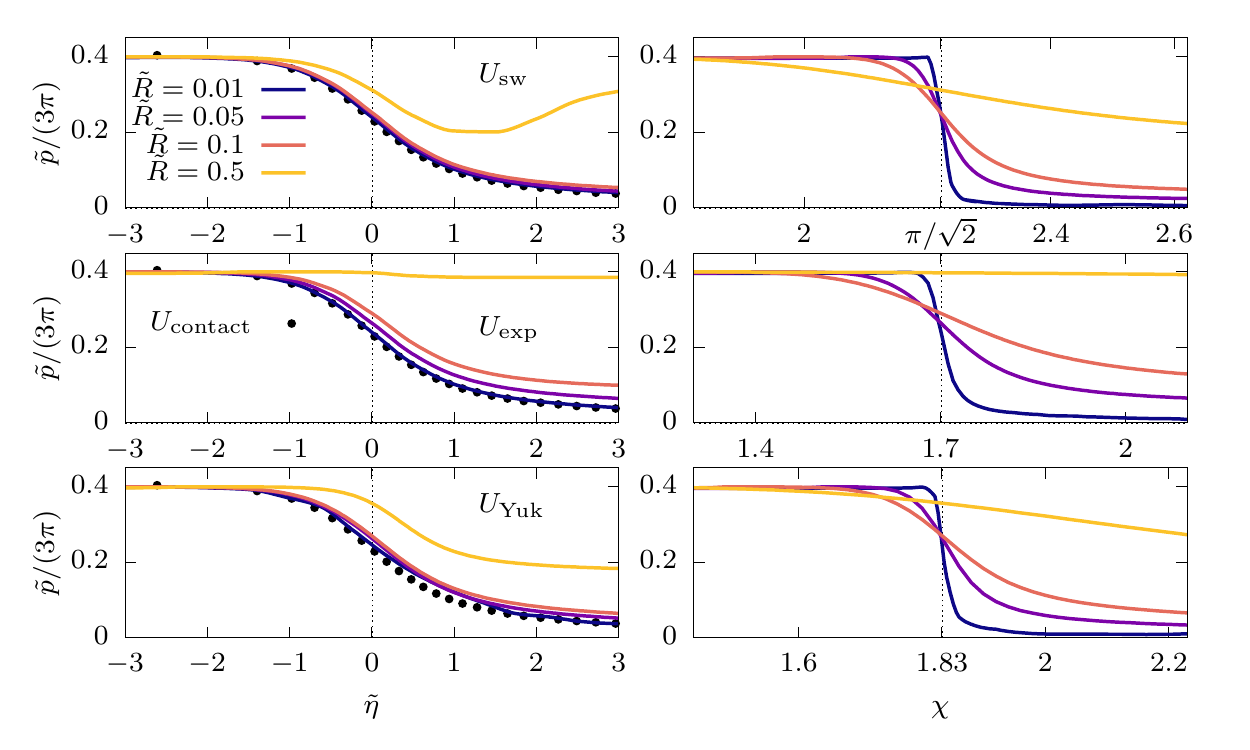} 
\caption{\footnotesize (Color online) Dimensionless pressure $\tilde p$ as a function of the inverse of the $s-$wave scattering length $\tilde \eta$ (left panels) and the dimensionless variable $\chi$ (right panels), for the square well $U_{\mathrm{sw}}$, exponential $U_{\mathrm{exp}}$, and Yukawa potentials $U_{\mathrm{Yuk}}$. The dotted line in the left panels is the contact approximation result.  Different curves in each panel correspond to different values of the potential range $\tilde R$. The vertical dotted line in the right panels indicate the values of $\chi$ at resonance for the different potentials.}
\label{pressure}
\end{figure}

In this section we analyze the thermodynamics along the crossover for the model potentials described in Eqs. (\ref{potentials}). In particular, we concentrate in determining the chemical potential $\tilde \mu$, the pressure $\tilde p$ and the energy gap $\tilde \Delta_{\mathbf{k}}$, as functions of the parameters $\tilde R$ and $\tilde V_0$. $\tilde \Delta_{\mathbf{k}}$ and $\tilde \mu$ were directly obtained by numerically solving the coupled system (\ref{SE-Delta-N2}), while the pressure $\tilde p$ was obtained from the grand potential function, recalling that $\Omega= -pV$, with $V$ the volume of the sample, see Eq.(\ref{Omega}). In Figs. \ref{gap}, \ref{mu} and \ref{pressure} we plot the quantities $\tilde \Delta_{\mathbf{k}}$, $\tilde \mu$ and $\tilde p$, respectively, as a function of both, 
the inverse of the $s-$wave scattering length $\tilde \eta$ (panels on the left), and the variable $\chi$ (panels on the right) for each interatomic potential. As indicated in the figures, we illustrate the behavior of those thermodynamic quantities for several values of the potential range $\tilde R =$ 0.01, 0.05, 0.1 and 0.5.  The dotted curves in all the panels on the left correspond to the contact interaction potential \cite{Leggett}. 

Several observations regarding the behavior of the thermodynamic quantities $\tilde \Delta_{\mathrm {max}}$, $\tilde \mu$ and $\tilde p$ follow from their corresponding figures. First, by looking at the left panels, where the thermodynamic variables are expressed as a function of $\tilde \eta$, one finds that 
for the smallest value of $\tilde R$ considered, $\tilde R = 0.01$, the standard result of the contact potential is recovered for all thermodynamic quantities and for all potentials. As $\tilde R$ is increased, while in the the BCS regime, $\tilde \eta < 0$, the agreement with the contact approximation remains, it is not true for the BEC side, $\tilde \eta > 0$. We point out here that we have included the case $\tilde R=0.5$, which is not very realistic since it corresponds to a very dense gas with an interatomic range nearly half of the mean interparticle separation, to make evident that significant deviations occur from the dilute case, estimated as initiating in $\tilde R \approx 0.1$. A very important question concerns the validity of the universality hypothesis \cite{Ho} at resonance, or unitarity, $\tilde \eta = 0$, which states that thermodynamic variables take on universal values at such a point. Although in the left panels one can assess whether the hypothesis holds or not, we find that the panels on the right, expressed in terms of the dimensionless variable $\chi$, on which the thermodynamic variables explicitly depend, its elucidation is certainly much clearer. That is, in agreement with the results shown of figure \ref{Fig1}, one should expect to find that at resonance, for a given model of interaction, all the observables should show a similar behavior at the resonance value of $\chi$. Indeed, we observe that, although universality strictly speaking does not hold, as evidently specified by the case $\tilde R=0.5$, it is restored as the gas becomes more diluted, namely, as $\tilde R \to 0$, independently of the potential. As one can see from the figures, the exponential potential shows a slightly disagreement for $\tilde R = 0.05$ and 0.1, but as the others, for $\tilde R = 0.01$, universality holds. These results certainly validate the bounds of the universality of thermodynamics at resonance. We return to this point in the final section of this work.
 The further gain in using the variable $\chi$, for finite range interactions, is that as the BEC side is entered, the differences in the thermodynamic variables are clearly shown even for very small values of the range $\tilde R$. 

\section{Bound molecules and Cooper pairs for finite range potentials}

One of the most remarkable signatures that have allowed to track the reversible smooth transition from the BCS superfluid to a BEC state in ultracold atomic experiments, is the presence of weakly-bound Cooper pairs and strongly-bound molecular dimers respectively. In the laboratory these pairs are usually created by means of a magnetically tunable resonance and have a definite and controllable binding energy.  Spectroscopic measurements of the pairing energy have confirmed the crossover from a two-body molecular pairing regime to the many body dominated BCS regime \cite{Chen,Cheng,Bartenstein}. In this work we also address the mentioned smooth transition by analyzing the nature of the pair functions in the whole range of the $s-$wave scattering length. Guided by the results of the previous section, we study pair functions for values of the scattering length well above the resonance, $a \rightarrow \pm \infty$. In particular, as shown in Ref. \cite{Ketterle}, the pair wavefunctions can be determined from the Fourier transform of the expectation value of  the creation pair operator $\hat a_{\mathbf{k}\uparrow}^\dagger a_{-\mathbf{k}\downarrow}^\dagger$, yielding,
\begin{equation}
\psi(r)= \frac{1}{V} \sum_{\mathbf{k}} u_{\mathbf{k}} v_{\mathbf{k}} e^{i \mathbf{k} \cdot \mathbf{r}},
\end{equation}
with the product $u_{\mathbf{k}}v_{\mathbf{k}}=\Delta_\mathbf{k}/2 E_{\mathbf{k}}$. The pair functions $\psi(r)$ can be shown to be real.

\begin{figure}[h!]
\includegraphics[width=0.7\textwidth]{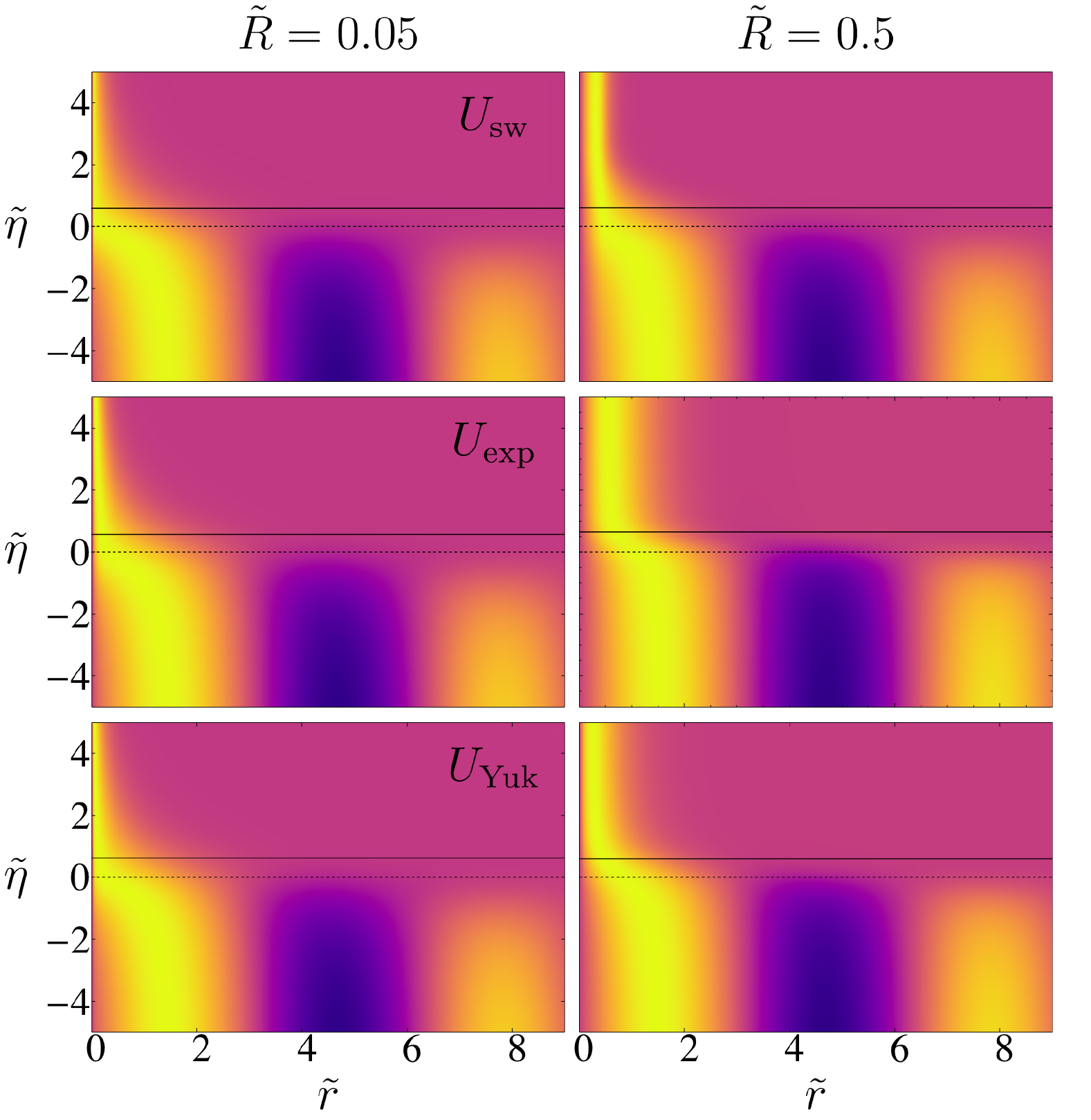}
\caption{\footnotesize (Color online) Density plot of the normalized pair wavefunctions ($\tilde\psi$) associated to different model potentials. Panels on the left and right correspond to potential ranges $\tilde R=0.05$ and $\tilde R=0.5$. Continuous and dashed lines indicate  $\mu=0$ and $\eta=0$ respectively. The brightest areas (yellow) correspond to $\tilde\psi=1$ interpolating to the darkest areas (purple) corresponding to $\tilde\psi=-1$, with $\tilde\psi=\tilde r\psi/\max(\tilde r\psi)$. }
 \label{pairF}
\end{figure}
In Fig. \ref{pairF} we plot the pair functions in a density-plot style to properly appreciate the behavior in the BCS and BEC regimes. These functions are normalized to their maximum value. Panels on the left and right correspond to the potential ranges extremes analyzed, $\tilde R=0.05$ and $\tilde R=0.5$. The continuous and dashed lines indicate the states $\tilde \mu=0$ and $\tilde \eta=0$, respectively. As one sees from Fig. \ref{pairF}, in the deep BCS and BEC regimes the pair functions clearly show the oscillatory and exponentially localized forms characteristic of the Cooper pairs and bound states respectively. While in the BEC regime one observes that the exponential decay does depend not only on the value of $\tilde R$, but it is also different for the different potentials, it is quite remarkable that the oscillatory behavior in the BCS side is essentially independent of the interatomic potential, with a weak dependence regarding the slow long range pair decay. That is, all cases show the same oscillatory behavior that can be fitted quite well as $\sin \tilde r = \sin k_F r$, indicating in turn that the oscillation wavelength is essentially the mean atomic separation $\overline d \sim 1/k_F$, regardless of the potential. 
An additional observation emerging from the pairing wave function behavior, close to the resonance is that, in agreement with the results of the previous section, and as stated by Leggett \cite{Leggett}, the smooth transition between BCS and BEC states does not seem to occur at the resonance $\eta = 0$ but it is spread out all the way to the state where the chemical potential changes its value from positive to negative.

\section{Final Remarks}

The present investigation was focused on the $T=0$ BCS-BEC crossover analysis of a two component mixture of fermions interacting through a finite range potential. In particular, working within the mean field approach, we considered three different model potentials, defined in terms of two independent parameters, the range $R$ and the potential depth $V_0$. Since pairing along the crossover is directly related with the interacting potential, we first studied the two-body physics described in terms of  the $s-$wave scattering length showing that the occurrence of a potential resonance, that is the emergence of a bound state, takes place at a fixed value of the variable $\chi=\tilde R \sqrt{\tilde V_0}$ for each potential. 

From the analysis of the thermodynamics at $T=0$ for the square well, exponential, and Yukawa potentials we arrived to one of the main results of this manuscript, that is, the non-universal character of the BCS-BEC transition, and its restoration. This is manifested in the dependence or not of the transition on a unified value of either the $s-$wave scattering length $a$, or the variable $\chi$, that encodes the information of $R$ and $V_0$. This result comes directly from the fact that the potentials here considered depend on two parameters, and can be explained as follows. According with the so called unitary limit, when $a$ diverges for a certain value of the potential depth $V_0$, this divergence overwhelms any other physical length that participates in the thermodynamics of the system, or alternatively, it indicates that the systems losses its length scales \cite{Ho}. Hence, not only $a$ but all the other lengths should not play any significant role on the thermodynamics. However, the thermodynamics of a system with finite range potential does not explicitly depend on the scattering length but, as in the cases here studied, on the two parameters $V_0$ and $R$ of the potential. This, as we have already discussed, yields two appropriate dimensionless parameters, $\chi = (2 m R^2 V_0/\hbar^2)^{1/2}$ and $\tilde R = k_F R$. Hence, at resonance, or unitarity, $\tilde \eta = 0$, $\chi$ takes its universal value for the given potential, but the thermodynamics still depends on $\tilde R$, thus indicating a non-universal behavior. It is therefore clear that as $\tilde R \to 0$, which is the dilute limit where ultimately the theory has its best performance, indeed, the potential range becomes irrelevant and, as a consequence, universal thermodynamics is restored. The other relevant results of our investigation are, first,  that the behavior of the pair functions allows us to distinguish that a smooth BCS-BEC crossover occurs between the change of signs of the scattering length and the chemical potential, with a clear slow decay and oscillations of purely thermodynamic origin in the BCS side, while in the BEC regime with an exponential decay depending both on thermodynamic variables as well as on the details of the interatomic potential. 

Our results can be relevant in view of both experiments performed with cold atoms in optical cavities, where tunable-range effective atomic interactions can be engineered, or in systems belonging to high energy physics where nuclear interactions can be properly modeled by specific potentials.

\acknowledgements{
E.N acknowledges financial support from a CONACYT scholarship. This work was partially funded by grants IN111516, IN105217 DGAPA (UNAM) and CONACYT 255573.
}

\end{document}